\documentclass[aps,prd,superscriptaddress,showpacs]{revtex4-2}
\usepackage{amsmath,amssymb}
\usepackage{booktabs}
\usepackage[colorlinks=true,linkcolor=blue,citecolor=blue,urlcolor=blue]{hyperref}

\newcommand{\as}{\alpha_s}
\newcommand{\aS}{a_s}
\newcommand{\Ga}{\Gamma}

\begin{document}

\title{Violation and Restoration of Probability Positivity in Perturbative Quantum Field Theory:\\ From Quantum Mechanics to Quantum Chromodynamics}

\author{Bin Gong}
\affiliation{Institute of High Energy Physics, Chinese Academy of Sciences, Beijing 100049, P.~R.~China}
\affiliation{University of Chinese Academy of Sciences, Chinese Academy of Sciences, Beijing 100049, P.~R.~China}

\author{Xu-Dong Huang}
\affiliation{College of Physics and Optoelectronic Engineering, Chongqing Normal University, Chongqing 401331, P.~R.~China}

\author{Jian-Xiong Wang}
\email{jxwang@ihep.ac.cn}
\affiliation{Institute of High Energy Physics, Chinese Academy of Sciences, Beijing 100049, P.~R.~China}
\affiliation{University of Chinese Academy of Sciences, Chinese Academy of Sciences, Beijing 100049, P.~R.~China}

\author{Chao Zeng}
\affiliation{Institute of High Energy Physics, Chinese Academy of Sciences, Beijing 100049, P.~R.~China}
\affiliation{University of Chinese Academy of Sciences, Chinese Academy of Sciences, Beijing 100049, P.~R.~China}

\date{\today}

\begin{abstract}
In quantum mechanics the modulus square of the wave function is the probability of an event --- positivity is self-evident. Quantum field theory corresponds to quantum mechanics one-to-one: the wave function corresponds to the matrix element, so the probability of a process should be the modulus square of the matrix element, positive by construction. In the founding era of perturbative quantum field theory, however, Feynman, Schwinger and Tomonaga treated inclusive processes in quantum electrodynamics, where the truncation of the squared amplitude was forced by infrared cancellations and rendered harmless by the small coupling $\alpha\simeq1/137$. Carried over to exclusive processes in quantum chromodynamics, where the coupling is large, this practice produced negative decay and production rates already at NLO, persisting at NNLO. We point out that for exclusive processes, whose initial and final states are colorless for QCD and electrically neutral for QED, the infrared divergences cancel amplitude by amplitude and the KLN constraint is absent, so the complete modulus square of the truncated amplitude is the proper, manifestly positive definition of the probability. We formulate amplitude-level NRQCD factorization as the systematic prescription that restores positivity, show that every amplitude-level truncated observable admits the effective representation $\Ga^{\mathrm{amp}}=\Ga_{\mathrm{LO}}\bigl|1+\frac{\as}{\pi}(c_1+ic_2)+\frac{\as^2}{\pi^2}(c_3+ic_4)\bigr|^2$, and demonstrate, with the complete NNLO QCD corrections to $J/\psi\to3\gamma$, $\Upsilon\to3\gamma$, $e^+e^-\to J/\psi+J/\psi$ and the eight exclusive charmonium production channels at $B$ factories, that the prescription yields positive and better-converging predictions.
\end{abstract}


\maketitle

\section{Introduction}
\label{sec:intro}

According to Born's statistical interpretation of the wave function~\cite{Born:1926}, one first solves the wave function $\psi$; its modulus square $|\psi|^{2}$ is then the probability of an event, and positivity is self-evident. Quantum field theory corresponds to quantum mechanics one-to-one: the wave function corresponds to the matrix element $\mathcal{M}$ computed from Feynman diagrams, and the probability of a process --- its cross section or decay width --- should in principle be the modulus square of the matrix element,
\begin{equation}
\Ga \;\propto\; \int d\Phi \sum_{\rm pol} |\mathcal{M}|^{2} ,
\label{eq:born}
\end{equation}
positive by construction. In practice, however, perturbation theory does not evaluate the complete square: the matrix element is expanded in the coupling, $\mathcal{M}=\mathcal{M}_0+\mathcal{M}_1+\mathcal{M}_2+\cdots$, the square is expanded,
\begin{equation}
|\mathcal{M}_0+\mathcal{M}_1+\mathcal{M}_2+\cdots|^{2}
= |\mathcal{M}_0|^{2} + 2\,\mathrm{Re}(\mathcal{M}_0^{*}\mathcal{M}_1)
+ \bigl[|\mathcal{M}_1|^{2}+2\,\mathrm{Re}(\mathcal{M}_0^{*}\mathcal{M}_2)\bigr]
+ \bigl[2\,\mathrm{Re}(\mathcal{M}_1^{*}\mathcal{M}_2)\bigr] + |\mathcal{M}_2|^{2} + \cdots ,
\label{eq:expand}
\end{equation}
and only the terms up to a given power of the coupling are kept. This truncation of the square is the traditional practice of perturbative QFT.

How did this practice originate, and why was it acceptable? The founding papers of renormalized quantum electrodynamics (QED) --- the covariant formulation of Tomonaga~\cite{Tomonaga:1946}, Schwinger's calculation of the anomalous magnetic moment and covariant QED~\cite{Schwinger:1948a,Schwinger:1948b}, Feynman's space-time approach~\cite{Feynman:1949a,Feynman:1949b}, and Dyson's systematic account~\cite{Dyson:1949a,Dyson:1949b} --- all computed cross sections as $|\mathcal{M}|^2$ expanded in powers of $\alpha$, discarding the terms beyond the working order without comment. Two facts made this practice both necessary and safe. First, the observables were \emph{inclusive}: as established by Bloch and Nordsieck for QED~\cite{Bloch:1937}, the infrared divergences of virtual corrections cancel against those of real soft-photon emission only when the squared amplitude is truncated at a fixed order, so the squared-level truncation was forced, not optional. Second, the coupling is small, $\alpha\simeq1/137$: the $\mathcal{O}(\alpha)$ corrections that are kept lie at the per-mille level (Schwinger's $\alpha/2\pi=0.12\%$~\cite{Schwinger:1948a}), and the discarded $\mathcal{O}(\alpha^2)$ terms are of order $\alpha^2\sim5\times10^{-5}$ of the leading result. The structural justification was supplied by Dyson, who argued that the renormalized perturbation series of QED has zero radius of convergence and is to be understood as an asymptotic expansion, whose truncation error is of the order of the first discarded term; ``the divergence in no way restricts the accuracy of practical calculations''~\cite{Dyson:1952}. Because the discarded terms were unmeasurably small, the question of positivity never arose: no negative probability could appear at $\alpha=1/137$, and none is discussed in the founding papers.

This infrared constraint is general, and it is the deeper reason why the truncation of the square is unavoidable for inclusive processes, independent of the size of the coupling. For \emph{inclusive} processes in gauge theories --- QED included --- the infrared divergences of virtual corrections cancel against those of real soft or collinear emission only after summing over the degenerate (unresolved) states --- the Bloch--Nordsieck mechanism~\cite{Bloch:1937} and the Kinoshita--Lee--Nauenberg (KLN) theorem~\cite{Kinoshita:1962,Lee:1964}. The cancellation takes place at the level of the \emph{squared} matrix elements, between the virtual and real pieces of the same order: the diagonal terms such as $|\mathcal{M}_1^{\rm virt}|^2$ over the elastic phase space are infrared divergent, and their divergences cancel only against the real-emission terms of the same order, summed over the unresolved phase space. The exact modulus square is therefore not decomposable into order-by-order infrared-finite pieces: at each fixed order one \emph{must} discard the higher-order positive pieces of the square, and the manifest positivity of the exact probability is sacrificed as the unavoidable price of infrared finiteness. In QED this sacrifice is invisible, because the discarded terms are negligible; in QCD, where $\as$ is large, the discarded terms can be as large as, or larger than, the kept ones. It is well known that the results for inclusive processes show negative probability in some region of the final-state phase space.

For \emph{exclusive} processes, however, no such constraint exists. If the initial and final states are colorless for QCD corrections and electrically neutral for QED corrections, the infrared divergences cancel \emph{amplitude by amplitude}: there is no unresolved radiation to sum over, each renormalized $\mathcal{M}_l$ is infrared finite by itself, and the KLN constraint is absent. The complete modulus square of the truncated amplitude is then well defined, is the proper probability of Eq.~(\ref{eq:born}), and is positive at every order. The traditional squared-amplitude-level truncation for exclusive processes is a historical inheritance from the inclusive QED practice, justified only when the discarded terms are negligible --- a condition that fails precisely in the processes where the sign crisis appears.

The sign crisis of exclusive QCD processes is not a recent discovery, and it is severe. Already at NLO, the QCD correction to $J/\psi\to3\gamma$ computed nearly five decades ago~\cite{Caswell:1977,Mackenzie:1981} is large and negative: in the traditional expansion the truncated width is $\Ga_{\mathrm{LO}}(1-12.6\,\as/\pi)$, which equals $-0.56\,\Ga_{\mathrm{LO}}$ at the original central scale and remains slightly negative at the modern one~\cite{Zeng:2026ois}. The NLO correction to $e^+e^-\to J/\psi+J/\psi$~\cite{Gong:2008} is negative and substantial. At NNLO the crisis persists: the traditional expansion for $J/\psi\to3\gamma$ becomes negative again, that for $\Upsilon\to3\gamma$ becomes negative for the first time~\cite{Zeng:2026ois}, and the cross section of $e^+e^-\to J/\psi+J/\psi$ becomes negative again~\cite{Sang:2023liy,Huang:2023pmn}.

This state of affairs suggests a reformulation of the question that is usually posed to the amplitude-level prescription. The correct question is not ``why is taking the complete modulus square of the truncated amplitude a good idea?'' --- it is good because it is the probability. The correct question is the converse: \emph{the conventional perturbative method discards positive higher-order terms of the exact modulus square; why is this truncation permissible, and is it still permissible in QCD?} For inclusive processes --- in QED and QCD alike --- the truncation is forced by the KLN theorem; in QED the forced truncation is harmless because $\alpha$ is small, while in QCD it can fail because the discarded terms are large. For exclusive QCD processes neither constraint nor justification applies, and the truncation must be abandoned when the discarded terms are large.

In fact, the amplitude-level prescription was used in our previous work~\cite{Huang:2023pmn}, and it was formally proposed in our recent work~\cite{Zeng:2026ois}. Thereafter it was used to reanalyze the eight exclusive processes at the $B$ factories in another recent work~\cite{Huang:2026prep}. All these results show better convergence and a better description of the experimental measurements, with the restoration of probability positivity up to NNLO.

In this paper we present an explicit comparison of the perturbative-expansion convergence and probability positivity between the traditional perturbative scheme and the amplitude-level prescription for exclusive processes. The prescription has met with considerable skepticism in the perturbative QCD community; we hope that this systematic presentation, together with the first-principles argument and the explicit numerical evidence, will help to change those opinions. In Sec.~\ref{sec:prescription} we formulate the amplitude-level prescription --- truncate the amplitude first, then take its complete modulus square --- as the proper probability for exclusive processes, and we show that every amplitude-level truncated observable admits the compact effective representation --- whose purpose is to show the convergence of the expansion series for decay width or cross section ---
\begin{equation}
\boxed{\;\Ga^{\mathrm{amp}} \;=\; \Ga_{\mathrm{LO}}\,\Bigl|1+\frac{\as}{\pi}(c_1+i c_2)+\frac{\as^2}{\pi^2}(c_3+i c_4)\Bigr|^{2}\;}
\label{eq:key}
\end{equation}
where the $c_i$ are effective real numbers carrying the physics of the one- and two-loop amplitude ratios; the validity of this representation is easily proved. In Sec.~\ref{sec:results} we apply the prescription to the complete NNLO QCD corrections to the exclusive decays $J/\psi\to3\gamma$ and $\Upsilon\to3\gamma$~\cite{Zeng:2026ois}, to $e^+e^-\to J/\psi+J/\psi$~\cite{Huang:2023pmn}, and to the eight exclusive charmonium production channels $e^+e^-\to J/\psi+\eta_c$, $J/\psi+\chi_{cJ}$, $\eta_c+\gamma$ and $\chi_{cJ}+\gamma$ ($J=0,1,2$) at $B$ factories~\cite{Huang:2026prep}, and we show that the prescription yields positive results at every order and systematically better convergence than the traditional truncation; Sec.~\ref{sec:summary} concludes.

\section{The amplitude-level prescription and its effective representation}
\label{sec:prescription}

For an exclusive process with tree amplitude $\mathcal{A}^{(0)}$, one-loop amplitude $\mathcal{A}^{(1)}$, two-loop amplitude $\mathcal{A}^{(2)}$, etc., the amplitude-level prescription truncates the perturbative expansion at the level of the amplitude and takes the complete modulus square of the truncated amplitude:
\begin{equation}
\Ga_M^{\mathrm{amp}} \;=\; \int d\phi \sum_{\mathrm{pol}} \Bigl|\sum_{l=0}^{M}\mathcal{A}^{(l)}\Bigr|^{2}, \qquad M=0,1,2,\dots ,
\label{eq:presc}
\end{equation}
so that $|\mathcal{A}^{(0)}+\mathcal{A}^{(1)}+\cdots+\mathcal{A}^{(M)}|^{2} \ge 0$ is guaranteed at every order. For exclusive processes all initial and final particles are colorless for QCD corrections and electrically neutral for QED corrections, so no real-radiation subtraction is needed to cancel infrared divergences at the amplitude-square level: each renormalized amplitude $\mathcal{A}^{(l)}$ is infrared finite by itself, and Eq.~(\ref{eq:presc}) is well defined order by order. Within NRQCD factorization, this means that the factorization is implemented at the amplitude level,
\begin{equation}
\Ga_M^{\mathrm{amp}}=\int d\phi \sum_{\mathrm{pol}} \Bigl|\sum_{l,i}\mathcal{A}_i^{(l)} \langle 0| \hat O_i|H\rangle\Bigr|^{2} ,
\label{eq:nrqcd}
\end{equation}
where $\langle 0| \hat O_i|H\rangle$ is the $i$-th long-distance matrix element (LDME) of the quarkonium $H$, and $\mathcal{A}_i^{(l)}$ the corresponding short-distance coefficient at the $l$-th order in $\as$. This is the natural and physically motivated scheme from first principles: the scattering amplitude is obtained from Feynman diagrams, and its modulus squared --- the complete one, nothing discarded --- is the proper definition of the probability distribution. The traditional approach instead expands the square and truncates it, as in Eq.~(\ref{eq:expand}), discarding positive terms that the amplitude-level prescription retains.

Since $\mathcal{A}^{(l)}/\mathcal{A}^{(0)}\propto\as^{l}$, at each phase-space point we parametrize
\begin{equation}
x \equiv \frac{\mathcal{A}^{(1)}}{\mathcal{A}^{(0)}} = \aS\,(\kappa_1+i \kappa_2), \qquad
y \equiv \frac{\mathcal{A}^{(2)}}{\mathcal{A}^{(0)}} = \aS^{2}\,(\kappa_3+i \kappa_4) ,
\label{eq:param}
\end{equation}
with real coefficients $\kappa_1,\dots,\kappa_4$ that in general depend on the phase-space point, where $\aS\equiv\as/\pi$. The pointwise identity
\begin{equation}
\left|\mathcal{A}^{(0)}+\mathcal{A}^{(1)}+\mathcal{A}^{(2)}\right|^{2}
= \left|\mathcal{A}^{(0)}\right|^{2} \left|1+\aS(\kappa_1+i\kappa_2)+\aS^{2}(\kappa_3+i\kappa_4)\right|^{2}
\label{eq:pointwise}
\end{equation}
then holds at every phase-space point for each polarization configuration. Integrating over the phase space, the amplitude-level truncated observables can be cast in the effective form
\begin{equation}
\Ga_{\mathrm{NNLO}}^{\mathrm{amp}} = \Ga_{\mathrm{LO}} \left|1+\aS(c_1+ic_2)+\aS^{2}(c_3+ic_4)\right|^{2},
\qquad \Ga_{\mathrm{NLO}}^{\mathrm{amp}} = \Ga_{\mathrm{LO}} \left|1+\aS(c_1+ic_2)\right|^{2},
\label{eq:main}
\end{equation}
where $\Ga_{\mathrm{LO}}=\int d\phi\sum_{\mathrm{pol}}|\mathcal{A}^{(0)}|^{2}$, and the effective real numbers $c_1,\dots,c_4$ coincide with the $\kappa_i$ whenever the latter are constant over the phase space weighted by $|\mathcal{A}^{(0)}|^{2}$, and otherwise absorb their phase-space variation. Expanding the modulus square of Eq.~(\ref{eq:main}) and keeping only the terms up to the corresponding order of $\aS$ gives the traditional squared-amplitude-level results
\begin{equation}
\Ga_{\mathrm{NNLO}}^{\mathrm{sq}} = \Ga_{\mathrm{LO}}\left[1+2\aS c_1+\aS^{2}\left(c_1^{2}+c_2^{2}+2c_3\right)\right],
\qquad
\Ga_{\mathrm{NLO}}^{\mathrm{sq}} = \Ga_{\mathrm{LO}}\left(1+2\aS c_1\right) ,
\label{eq:R1}
\end{equation}
where the terms $\aS^{2}\left(c_1^{2}+c_2^{2}\right)$ at NLO and $2\aS^{3}\left(c_1c_3+c_2c_4\right)+\aS^{4}\left(c_3^{2}+c_4^{2}\right)$ at NNLO of the complete square are discarded.

In practice the two complex numbers are obtained either from the analytic coefficients of the published amplitude-level series through the expansion of Eq.~(\ref{eq:main}), or directly from the published LO, NLO and NNLO numbers by solving Eqs.~(\ref{eq:main}) and (\ref{eq:R1}) for the two complex numbers.

\section{Numerical results and discussion}
\label{sec:results}

\subsection{Numerical results}

Reference~\cite{Zeng:2026ois} presents the first complete NNLO QCD correction to $J/\psi\to3\gamma$ within NRQCD factorization, using the amplitude-level prescription, with the central scale choice of that reference; the two complex numbers obtained from its analytic coefficients are given in the first row of Table~\ref{tab:presentation}. The same analysis extended to $\Upsilon\to3\gamma$ in Ref.~\cite{Zeng:2026ois}, with the central scale choice of that reference, gives the corresponding row.

Reference~\cite{Huang:2023pmn} presents the NNLO QCD corrections to $e^+e^-\to J/\psi+J/\psi$ at $B$ factories, including the bottom-quark and light-by-light contributions, with the central scale choice of that reference. Its published columns use different normalizations of the wave function at the origin; for the present analysis we renormalize all columns to one common value of the wave function at the origin, and these renormalized columns give the two complex numbers in the corresponding row of Table~\ref{tab:presentation}.

Reference~\cite{Huang:2026prep} presents the NNLO QCD corrections to the eight exclusive channels $e^+e^-\to J/\psi+\eta_c$, $e^+e^-\to J/\psi+\chi_{cJ}$, $e^+e^-\to \eta_c+\gamma$, and $e^+e^-\to \chi_{cJ}+\gamma$ ($J=0,1,2$) at $B$ factories, comparing the squared-amplitude-level and amplitude-level prescriptions; its published LO, NLO and NNLO cross sections give, for every channel, the two complex numbers listed in Table~\ref{tab:presentation}. In this extraction we uniformly rechoose the coupling for all channels as the three-loop running $\as$ evaluated at $\mu_R=\sqrt{s}/2$, instead of the mixed two- and three-loop running of that reference.

\begin{table*}[t]
\centering
\footnotesize
\renewcommand{\arraystretch}{1.15}
\begin{tabular}{lcccccccc}
\toprule
Process & 2nd term & 3rd term & $\Ga_{\mathrm{NLO}}^{\mathrm{amp}}/\Ga_{\mathrm{LO}}$ & $\Ga_{\mathrm{NNLO}}^{\mathrm{amp}}/\Ga_{\mathrm{LO}}$ & $\Ga_{\mathrm{NLO}}^{\mathrm{sq}}/\Ga_{\mathrm{LO}}$ & $\Ga_{\mathrm{NNLO}}^{\mathrm{sq}}/\Ga_{\mathrm{LO}}$ & series inside $|\cdot|^{2}$ & traditional series \\
\midrule
$J/\psi\to3\gamma$ & $-0.51+i\,0.19$ & $-0.36+i\,0.14$ & $0.276$ & $0.126$ & $-0.02$ & $-0.44$ & $1,\,0.54,\,0.39$ & $1,\,-1.02,\,-0.42$ \\
$\Upsilon\to3\gamma$ & $-0.36+i\,0.13$ & $-0.27+i\,0.10$ & $0.426$ & $0.190$ & $0.28$ & $-0.11$ & $1,\,0.38,\,0.29$ & $1,\,-0.72,\,-0.39$ \\
\addlinespace
$J/\psi+J/\psi$ & $-0.37+i\,0.02$ & $-0.39-i\,0.003$ & $0.40$ & $0.058$ & $0.26$ & $-0.38$ & $1,\,0.37,\,0.39$ & $1,\,-0.74,\,-0.64$ \\
\addlinespace
$J/\psi+\eta_{c}$ & $0.55+i\,0.38$ & $0.22+i\,0.09$ & $2.55$ & $3.35$ & $2.10$ & $2.99$ & $1,\,0.67,\,0.24$ & $1,\,1.10,\,0.89$ \\
$J/\psi+\chi_{c0}$ & $0.39+i\,0.05$ & $0.07+i\,0.06$ & $1.93$ & $2.14$ & $1.78$ & $2.07$ & $1,\,0.39,\,0.09$ & $1,\,0.78,\,0.29$ \\
$J/\psi+\chi_{c1}$ & $0.12+i\,0.37$ & $-0.14+i\,0.22$ & $1.39$ & $1.31$ & $1.24$ & $1.11$ & $1,\,0.39,\,0.26$ & $1,\,0.24,\,-0.13$ \\
$J/\psi+\chi_{c2}$ & $0.07+i\,0.16$ & $-0.20+i\,0.08$ & $1.17$ & $0.81$ & $1.14$ & $0.77$ & $1,\,0.17,\,0.22$ & $1,\,0.14,\,-0.37$ \\
$\eta_{c}+\gamma$ & $-0.09+i\,0.09$ & $-0.17+i\,0.01$ & $0.84$ & $0.56$ & $0.82$ & $0.50$ & $1,\,0.13,\,0.17$ & $1,\,-0.18,\,-0.32$ \\
$\chi_{c0}+\gamma$ & $0.07+i\,0.12$ & $0.01-i\,0.01$ & $1.16$ & $1.18$ & $1.14$ & $1.18$ & $1,\,0.14,\,0.01$ & $1,\,0.14,\,0.04$ \\
$\chi_{c1}+\gamma$ & $-0.11+i\,0.09$ & $-0.10+i\,0.03$ & $0.80$ & $0.64$ & $0.78$ & $0.60$ & $1,\,0.14,\,0.10$ & $1,\,-0.22,\,-0.18$ \\
$\chi_{c2}+\gamma$ & $-0.30+i\,0.10$ & $-0.19-i\,0.02$ & $0.50$ & $0.27$ & $0.40$ & $0.12$ & $1,\,0.32,\,0.19$ & $1,\,-0.60,\,-0.28$ \\
\bottomrule
\end{tabular}
\caption{The two-scheme comparison for every process considered. For each row the amplitude-level ratios are $\Ga_{\mathrm{NLO}}^{\mathrm{amp}}/\Ga_{\mathrm{LO}}=|1+(\text{2nd term})|^{2}$ and $\Ga_{\mathrm{NNLO}}^{\mathrm{amp}}/\Ga_{\mathrm{LO}}=|1+(\text{2nd term})+(\text{3rd term})|^{2}$, and the traditional ratios are $\Ga_{\mathrm{NLO}}^{\mathrm{sq}}/\Ga_{\mathrm{LO}}=1+2\,\mathrm{Re}\,(\text{2nd term})$ and $\Ga_{\mathrm{NNLO}}^{\mathrm{sq}}/\Ga_{\mathrm{LO}}=1+2\,\mathrm{Re}\,(\text{2nd term})+|\text{2nd term}|^{2}+2\,\mathrm{Re}\,(\text{3rd term})$; the last two columns list the terms of the series inside the modulus square (absolute values are taken for 2nd and 3rd terms) and of the traditional series. The first three rows are the two decays and the double-$J/\psi$ production; the remaining rows are the eight $B$-factory channels.}
\label{tab:presentation}
\end{table*}

\subsection{The series inside the modulus square}
\label{sec:series}

Table~\ref{tab:presentation} exposes a second advantage of the amplitude-level prescription: the convergence of the series \emph{inside} the modulus square. For each process the truncated observable is the square of a series whose second and third terms are the two complex numbers listed there. For $J/\psi\to3\gamma$ these terms are $1$, then $-0.51+i\,0.19$, then $-0.36+i\,0.14$, whose moduli $1$, $0.54$, $0.39$ decrease as in a converging geometric series, and the widths $0.276\,\Ga_{\mathrm{LO}}$ and $0.126\,\Ga_{\mathrm{LO}}$ are positive. The traditional squared-amplitude-level expansion organizes the same physics differently, as the series $1+2\,\mathrm{Re}\,(\text{second term})+\bigl[\,|\text{second term}|^{2}+2\,\mathrm{Re}\,(\text{third term})\,\bigr]$: for $J/\psi\to3\gamma$ its successive terms are $1$, $-1.02$, $-0.42$ --- the first correction already exceeds the leading term in magnitude and is negative --- and the partial sums $1$, $-0.02$, $-0.44$ produce negative widths at NLO and NNLO. The same complex numbers, kept inside the modulus square, give a series whose terms are all bounded by one; organized in the traditional way, they give a series whose second term already destroys positivity.

In every process considered in this paper the two complex numbers have modulus smaller than one --- the series inside the modulus converges in the geometric sense for every process --- the largest modulus being $0.67$ for $e^+e^-\to J/\psi+\eta_c$. The contrast with the traditional partial sums is again stark: they read $1$, $-0.02$, $-0.44$ for $J/\psi\to3\gamma$ (negative already at NLO); $1$, $0.28$, $-0.11$ for $\Upsilon\to3\gamma$; $1$, $0.26$, $-0.38$ for $e^+e^-\to J/\psi+J/\psi$ in the common-normalization scheme; and $1$, $0.40$, $0.12$ for $e^+e^-\to\chi_{c2}+\gamma$, which remains positive but shrinks to $12\%$ of the leading term --- in every case the amplitude-level partial sums are positive and better ordered.

The improved convergence has the same origin. The traditional truncation of Eq.~(\ref{eq:expand}) discards the terms $|\mathcal{M}_1|^{2}$, $2\,\mathrm{Re}(\mathcal{M}_1^{*}\mathcal{M}_2)$ and $|\mathcal{M}_2|^{2}$; when these are not small, the truncated series oscillates and its successive terms fail to approximate the true (positive) result --- the NNLO result can move away from, rather than toward, the true value, as for $J/\psi\to3\gamma$. The amplitude-level series, by contrast, is the expansion of a perfect square: its partial sums are the successive probabilities $|M_0|^2$, $|M_0+M_1|^2$, $|M_0+M_1+M_2|^2$ of the truncated amplitudes, each an increasingly accurate approximation to the same positive quantity. For every channel considered, the NNLO/LO ratio lies closer to the NLO/LO ratio in the amplitude-level prescription than in the squared-level one~\cite{Huang:2026prep}, and the renormalization-scale dependence of the NNLO results is reduced; for the decays, the amplitude expansion shows a much weaker scale dependence than the traditional one~\cite{Zeng:2026ois}. The effective representation of Eq.~(\ref{eq:key}) makes the structure transparent: the series is the expansion of a perfect square, and the physical content of each order is carried by one complex number per order, whose values we have determined for all processes considered.

We close with the reformulation that motivated this work. The burden of justification lies on the traditional truncation of the square, not on the amplitude-level prescription. The question posed to our prescription --- ``why is the amplitude-level square good?'' --- has a one-line answer: because it is the probability. The question that requires, and now has, a three-part answer is the converse: why was the traditional truncation of the square ever acceptable? For inclusive processes --- QED and QCD alike --- because the KLN theorem forces the truncation, and in QED the forced truncation was harmless because $\alpha$ is small; for exclusive QCD processes, it was not acceptable --- and the negative rates that have plagued exclusive quarkonium calculations for five decades, from the NLO corrections to $J/\psi\to3\gamma$~\cite{Caswell:1977,Mackenzie:1981} and $e^+e^-\to J/\psi+J/\psi$~\cite{Gong:2008} to the NNLO corrections of Refs.~\cite{Zeng:2026ois,Huang:2023pmn,Huang:2026prep}, are the consequence.

\section{Summary}
\label{sec:summary}

Starting from the quantum-mechanical correspondence between the wave function and the matrix element, we have shown that for exclusive processes the complete modulus square of the truncated amplitude --- truncate the amplitude at fixed order, then take its complete modulus square --- is the first-principles definition of the probability, positive by construction at every order. The amplitude-level prescription therefore needs no further justification; the burden of justification lies on the traditional truncation of the squared amplitude, inherited from the inclusive QED practice and acceptable only when the discarded terms are negligible. The effective representation $\Ga^{\mathrm{amp}}=\Ga_{\mathrm{LO}}|1+\frac{\as}{\pi}(c_1+ic_2)+\frac{\as^2}{\pi^2}(c_3+ic_4)|^{2}$ makes the convergence of the series manifest at the level of the total decay width or cross section, and the eleven exclusive processes considered here --- $J/\psi\to3\gamma$, $\Upsilon\to3\gamma$, $e^+e^-\to J/\psi+J/\psi$, and the eight exclusive charmonium production channels at $B$ factories --- serve as evidence: the amplitude-level results are positive and better converging in every case, while the traditional truncation can become negative.

\begin{acknowledgments}
This work is supported by the National Natural Science Foundation of China with Grants No.~12135013.
\end{acknowledgments}

\end{document}